\documentclass[twocolumn,twoside,slac_two]{revtex4}
\usepackage{graphicx}
\usepackage{fancyhdr}
\usepackage{natbib}

\begin{document}

\title{On the high energy pulsar population detected by Fermi}

%

\author{G. A. Caliandro\footnote{andrea.caliandro@ieec.uab.es}}
\affiliation{Institut de Ciencies de l’Espai (IEEC-CSIC), Campus UAB, 08193 Barcelona, Spain}
\author{E. C. Ferrara}
\affiliation{NASA Goddard Space Flight Center, Greenbelt, MD 20771, USA}
\author{D. Parent}
\affiliation{CNRS/IN2P3, Centre d’´ Etudes Nucl´eaires Bordeaux Gradignan, UMR 5797,
Gradignan, 33175, France}
\author{R. W. Romani}
\affiliation{W. W. Hansen Experimental Physics Laboratory, Kavli Institute for Particle
Astrophysics and Cosmology, Department of Physics and SLAC National Accelerator
Laboratory, Stanford University, Stanford, CA 94305, USA}
\author{on behalf of the Fermi Large Area Telescope collaboration}
\affiliation{}

\begin{abstract}
The Large Area Telescope (LAT), Fermi's main instrument, is providing a new view of the local energetic pulsar population.
In addition to identifying a pulsar origin of a large fraction of the bright unidentified Galactic EGRET sources, the LAT
results provide a great opportunity to study a sizable population of high-energy pulsars.
Correlations of their physical properties, such as the trend of the luminosity versus the rotational energy loss rate, help
identify global features of the gamma-ray pulsar population. Several lines of evidence, including the light curve and
spectral features, suggest that gamma-ray emission from the brightest pulsars arises largely in the outer magnetosphere.
\end{abstract}

\maketitle

\thispagestyle{fancy}


\section{INTRODUCTION}
Since the launch of the Fermi satellite and during its first six months of data taking in survey mode the pulsar population discovered in the gamma-ray sky is grown as far as 46 objects. 
All their features and a first study of the new population are summarized in The First Fermi Large Area Telescope Catalog of Gamma-ray Pulsars \citep{psrCat}.
Fermi data were analyzed to search for pulsations using 762 contemporaneous ephemerides obtained from radio telescopes and 5 from X-ray telescopes. A group of 218
of these ephemerides come from a list of good candidate gamma-ray pulsars selected with spin-down power greater than $10^{34}$ erg s$^{-1}$. The rest of the ephemerides is a
sample of 544 pulsars from nearly the entire $P-\dot P$ plane, reducing in this way the possible bias of the LAT pulsar searches. In total the LAT detected 29 radio-selected pulsars, 5 of them are EGRET pulsars, and 8 are MSPs \citep{mspsr}.
The LAT has also discovered 16 gamma-selected pulsars \citep{blindpsr}. The discovery of these pulsars, as well as the determination of their timing ephemerides come directly from
LAT gamma-ray data. Including Geminga, the new population counts 17 gamma-selected pulsars. A campaign to search for their radio counterpart started soon and three
of them have already been found \citep{2009ApJ...705....1C}\citep{psrj1907}. However, with good probability some of them are true radio-quiet gamma-ray pulsars, as in the case of CTA1 \citep{cta1} and PSR
1836+5925, that were deeply searched for radio pulsations in the past.
With this large population of pulsars statistical studies on their features can be performed, that allow to better understand their physics and emission mechanisms.


\begin{figure}[b]
\includegraphics[width=0.9\columnwidth]{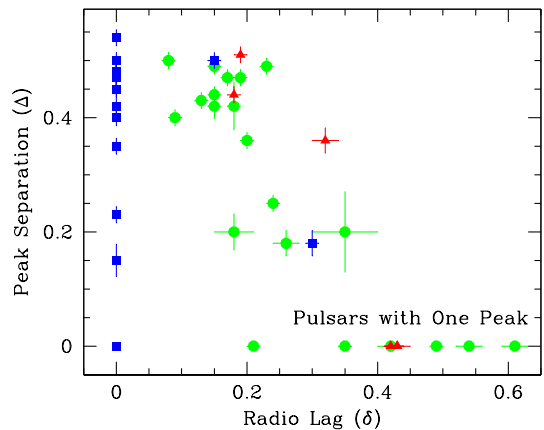}
\caption{Phase difference $\Delta$ between the gamma-ray peaks, versus
the phase lag $\delta$ between the main radio peak and the nearest gamma-ray
peak. Pulsars without a radio detection are plotted with $\delta=0$. With
present light curves we cannot generally measure $\Delta < 0.15$; objects
classified as single-peaked are plotted with $\Delta$=0. Two such
objects, both MSPs, are off the plot at $\delta>0.8$. 
Blue squares: gamma-ray-selected pulsars. 
Red triangles: millisecond gamma-ray pulsars. 
Green circles: all other radio loud gamma-ray pulsars.
\label{RadioSep}}
\end{figure}

\begin{figure}
\includegraphics[width=0.9\columnwidth]{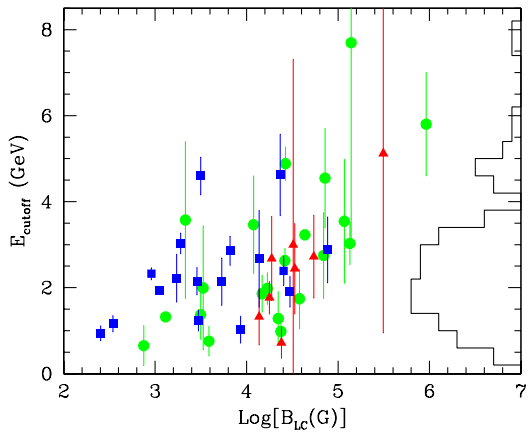}
\caption{Value of the exponential cutoff $E_{\rm cutoff}$ versus the magnetic field at the light cylinder, $B_{\rm LC}$.
The histogram of $E_{\rm cutoff}$ values is projected along the right-hand axis.
Blue squares: gamma-ray-selected pulsars.
Red triangles: millisecond gamma-ray pulsars. Green circles: all other radio loud gamma-ray pulsars. 
\label{EcvsBlc}}
\end{figure}

\section{TIMING AND SPECTRAL FEATURES}
Deep analysis were performed on all the pulsars detected by Fermi. Here we summarize and discuss their results. A detailed description of these analysis can been found in the first Fermi pulsar catalog \citep{psrCat}.

Long base ephemerides that cover the first six months of LAT sky survey were evaluated for each one of the detected pulsars. For the radio loud pulsars they are built joining the radio data of several radio telescopes. All the telescopes involved in this work are Parkes, Nan\c cay, Westerbork Synthesis radio telescopes, Arecibo, Green Bank Telescope, and the 76-m Lovell radio telescope at Jodrell Bank.
For the gamma-selected pulsars the timing solutions are evaluated using the LAT-data with a technique described in \citep{BlindTiming}.
The LAT observation is divided in several time windows of 10-20 days depending on the brightness of the source; the data in the windows are analyzed in order to define a pulse time of arrival (ToA) for each of them, that are finally used to build the time solution over the entire period of observation.

The light curves obtained folding the photons with the calculated ephemeris were analyzed and parametrized. 
As first, they have been classified for their peak multiplicity, as single or double peaked. In the latter case the distance in phase of the two gamma-ray peaks ($\Delta$) was measured.
Most of the radio loud gamma-ray pulsars have a radio profile as simple as a single peak, or it is always possible in general to clearly define a main radio peak. The phase delay of the first gamma-ray peak respect to the main radio peak ($\delta$) was evaluated.

Most of the pulsars show two
dominant, relatively sharp
peaks, suggesting the caustics
from the edge of a hollow cone.
When a single peak is seen, it
tends to be broader, suggesting
a tangential cut through an
emission cone. This picture is
realized in the Outer Gap and
the high altitude portion of the
Slot Gap models. As well, these
models predict the 
correlation observed in figure \ref{RadioSep} 
in the radio lag ($\delta$) versus the the gamma-ray peaks separation ($\Delta$).

\begin{figure}
\includegraphics[width=0.9\columnwidth]{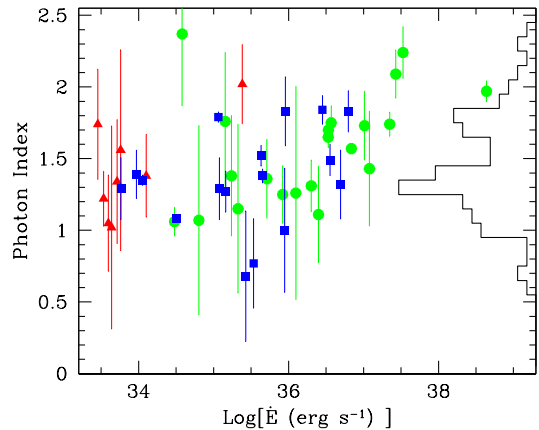}
\caption{Photon index $\Gamma$ versus the rotational energy loss rate, $\dot E$. 
The histogram of the photon indices is projected along the right-hand axis.
Blue squares: gamma-ray-selected pulsars.
Red triangles: millisecond gamma-ray pulsars. Green circles: all other radio loud gamma-ray pulsars. 
\label{IndexvsEdot}}
\end{figure}

\begin{figure*}
\centering
\includegraphics[width=0.9\textwidth]{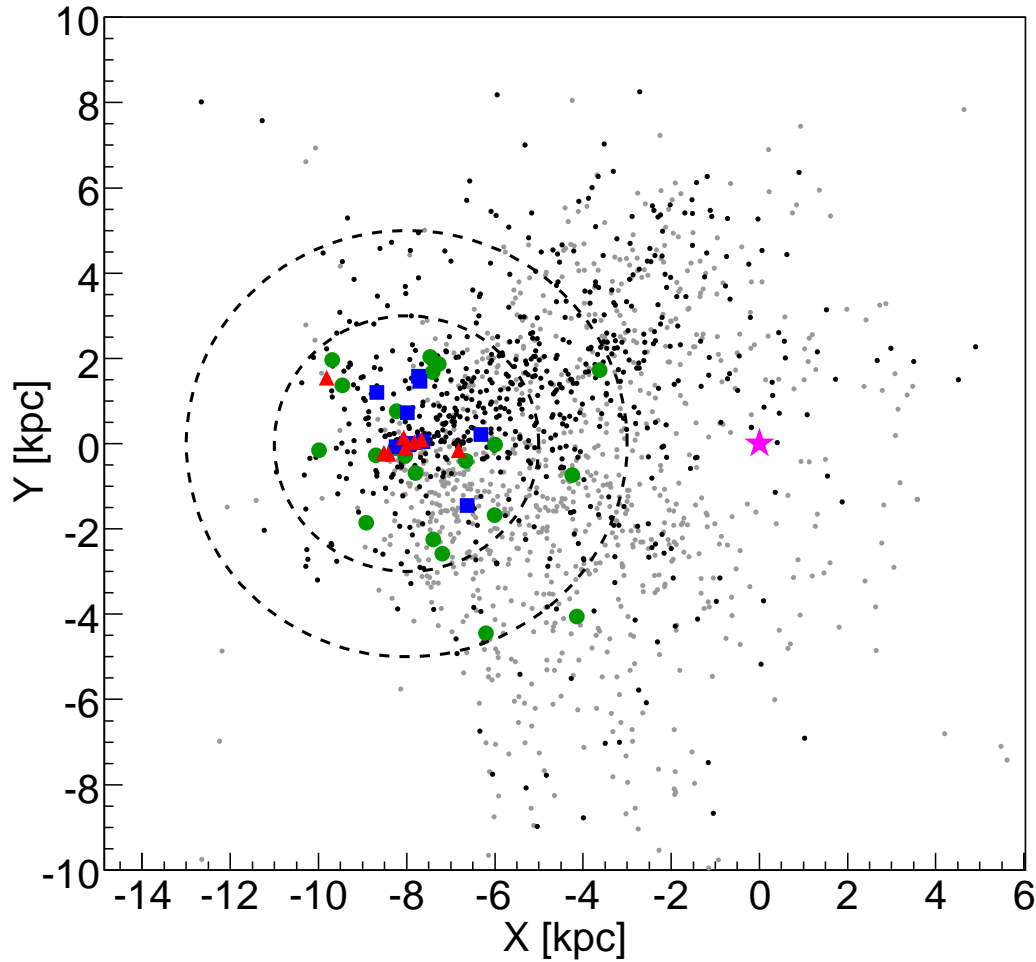}
\caption{Galactic plane pulsar distribution (polar view). The star represents the Galactic center. 
The two circles centered at the Earth's position have radii of 3 kpc and 5 kpc.  
For pulsars with different possible distances, the nearer values are used. 
Blue squares: gamma-ray-selected pulsars.
Red triangles: millisecond gamma-ray pulsars. Green circles: all other radio loud gamma-ray pulsars. 
Black dots: Pulsars for which gamma-ray pulsation searches were conducted using rotational ephemerides.
Gray dots: Known pulsars which were not searched for pulsations.
\label{SkyProjection}}
\end{figure*}

The spectrum energy of all the LAT gamma-ray pulsars is
well fitted by a power-law with an exponential cut-off. 

\begin{equation}
\frac{{\rm d} N}{{\rm d} E} = K E_{\rm GeV}^{-\Gamma}
                              \exp \left(- \frac{E}{E_c} \right)
\label{expcutoff}
\end{equation}

The significance of the exponential cut-off respect of a simply power-law model has been evaluated with
the likelihood ratio test. For some pulsars the significance is below 5$\sigma$, and the cut-off energy ($E_c$) 
is poorly determined for those few with the significance lower than 3$\sigma$.

In figure \ref{EcvsBlc} is plotted the cut-off energy versus the magnetic field at the light cylinder ($B_{LC}$). 

\begin{equation}
 B_{\rm LC} = \left( \frac{3I8\pi^4\dot{P}}{c^3P^5} \right)^{1/2} \approx 2.94 
 \times 10^{8}(\dot{P}P^{-5})^{1/2}\, {\rm G}.
\end{equation}

Along the right axis of figure \ref{EcvsBlc} is shown the histogram of the cut-off energies ($E_c$), 
that have a range of only
about a decade, from 1 to 10 GeV, and are weakly
correlated with the magnetic field at the light cylinder ($B_{LC}$).

The fact that this features are common to all the different
types of pulsars strongly implies that the gamma-ray
emission originates in similar locations in the
magnetosphere relative to the light cylinder. Such a
correlation of $E_c$ with $B_{LC}$ is actually expected in all outer
magnetosphere models where the gamma-ray emission
primarily comes from curvature radiation of electrons
whose acceleration is balanced by radiation losses \citep{Muslimov04} \citep{Zhang04} \citep{Hirotani08}
[12].

In figure \ref{IndexvsEdot} is plotted the spectral index versus the rotational energy loss rate ($\dot E$). As shown in the histogram along the right axis, the spectral indices are distributed around $\sim 1.5$ and there
is a general trend for the young pulsars to show a softer
spectrum at large $\dot E$. This may be indicative of higher pair
multiplicity, which would steepen the spectrum for the
more energetic pulsars, either by steepening the spectrum
of the curvature radiation-generating primary electrons
\citep{Romani96} or by inclusion of an additional soft spectral
component associated with robust pair formation \citep{Takata07} \citep{Harding08}.

\begin{figure*}
\centering
\includegraphics[width=0.9\textwidth]{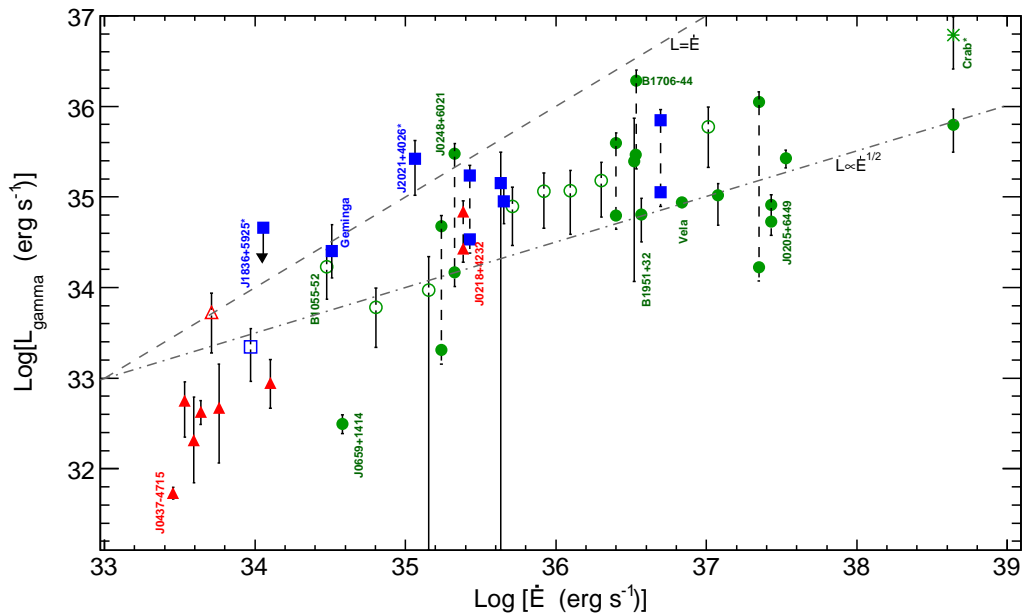}
\caption{Gamma-ray luminosity $L_\gamma$ versus the rotational energy loss rate $\dot E$. 
Dashed line: $L_\gamma$ equal to $\dot E$. 
Dot-dashed line: $L_\gamma$ proportional to the square root of $\dot E$. 
$L_\gamma$ is calculated using a beam correction factor $f_\Omega=1$ for all pulsars.
For the Crab is also plotted the total high energy luminosity, $L_{\rm tot} = L_X + L_\gamma$, indicated by *.
Several notable pulsars have been labeled.
Blue squares: gamma-ray-selected pulsars.
Red triangles: millisecond gamma-ray pulsars. Green circles: all other radio loud gamma-ray pulsars. 
Unfilled markers indicate pulsars for which only a DM-based distance estimate is available.
Pulsars with two distance estimates have two markers connected with dashed error bars.
\label{LEdot}}
\end{figure*}

\section{THE LOCAL POPULATION AND THE PULSAR LUMINOSITY}

To evaluate the Galactic pulsar distribution as well as their luminosity, a good knowledge of the
distances is needed. The annual trigonometric parallax method give the most precise estimate of
the distance, but unfortunately it is available only for few nearby pulsars. Among the other methods,
the most commonly used estimates the pulsar distance from the radio dispersion measure (DM)
coupled to an electron density distribution model, as the NE2001 \citep{Cordes2002}. The distances measured with this method have to be taken carefully and a conservative error of 30\% is assumed for the calculation of the pulsar luminosity.
When distances from different
methods disagree and no one is more convincing than the other, a distance range is assumed.

In figure \ref{SkyProjection} the pulsar distribution on the Galactic plane clearly shows that the LAT is detecting a local
population of gamma-ray pulsars, most of them not farther than 3 kpc (the inner circle in the figure). For the pulsars with an estimated distance range, only the closer edge of the range is drawn in the figure.

In order to give a complete description of the gamma-ray pulsar population, the possible association with sources in other wavelengths has been investigated.
At least 19 of the 46 LAT gamma-ray pulsars are positional associated with a PWN and/or SNR,
and 9 of them are associated with TeV sources. Those pulsars with both TeV and PWN
associations are typically young, with ages less than 20 kyr. There are at least 3 more pulsars
associated with TeV sources but not with PWNs or SNR.

Among the LAT gamma-ray pulsars, 17 were unidentified EGRET sources. This result improves the
discussions about the nature of the Galactic unidentified EGRET sources, as well as the discussion
on the pulsars emission models. Population synthesis studies have shown that the expected ratio
of radio-loud and radio-quiet pulsars strongly depends on the assumed emission model. The total
Galactic birthrate of energetic pulsars evaluated from the LAT sample is of ~1/50 yr, with gamma-selected
object representing half or more. The small ratio of radio-selected to gamma-selected
gamma-ray pulsars suggests that gamma-ray emission has an appreciably larger extent than the
radio beams, such as expected in the outer gap (OG) \citep{Romani1995} and slot-gap/two pole caustic (SG/TPC)
models \citep{Muslimov04} \citep{2003ApJ...598.1201D}.

Other constraints on pulsar models can come from the investigation of the relation between the pulsar luminosity ($L_{\gamma}$) and the rotational energy loss rate ($\dot E$).  In figure \ref{LEdot} is plotted this relation for the gamma ray pulsars detected by LAT.
The dashed line indicates $L_\gamma=\dot{E}$, while the dot-dashed line indicates $L_\gamma\propto\dot{E}^{1/2}$, where 
\begin{equation}
  L_\gamma \equiv 4\pi d^2 f_\Omega G_{100}.
\end{equation}

The  flux correction factor $f_\Omega$ \citep{Watters09} is 
model-dependent and depends on the magnetic inclination and observer angles $\alpha$ 
and  $\zeta$. 
Both the outer gap and slot gap models predict $f_\Omega\sim1$, 
in contrast to earlier use of $f_\Omega=1/4\pi\approx0.08$ \citep[in e.g.][]{Thompson94}, 
or $f_\Omega=0.5$ for MSPs \citep[in e.g.][]{Fierro95}. For simplicity, we use 
$f_\Omega=1$ throughout the paper, which presumably induces
an artificial spread in the quoted $L_\gamma$ values. 
However, it is the quadratic distance dependence for $L_\gamma$ that dominates
the uncertainty in $L_\gamma$
in nearly all cases. 

Gamma-rays dominate the total power $L_{\rm tot}$ radiated by most known high-energy pulsars, 
that is, $L_{\rm tot}\approx L_\gamma$. The Crab is a notable exception, 
with X-ray luminosity $L_X\sim10L_\gamma$. 
In Figure \ref{LEdot} we plot both $L_\gamma$ and $L_X + L_\gamma$. 
$L_X$  for $E<100$ MeV is taken from Figure 9 of \citet{Kuiper2001}.

Many models for the gamma-ray emitting gap predict that the gamma-ray
luminosity of the pulsars is proportional to the root square of the spin-down
power ($L_{\gamma} \propto \dot E$). This imply that the gamma-ray efficiency increases with
decreasing Ė, down to $\sim 10^{34}-10^{35}$ erg s$^{-1}$ where the gap saturates at large
efficiency. However the luminosity evolution of the LAT pulsars sample seems
more complicated.
At high $\dot E$ ($>10^{36.5}$ erg s$^{-1}$) the data seems to be in rough agreement with the
predicted trend. For $10^{35}<\dot E<10^{36.5}$ erg s$^{-1}$ $L_{\gamma}$ seem nearly constant, but the
uncertainties in this range are very large and could hide the true luminosity
evolution. In the range $10^{34}<\dot E<10^{35}$ erg s$^{-1}$ the gap saturation is expected to
occur in both Slot Gap ($\sim10^{35}$ erg s$^{-1}$) and Outer Gap models ($\sim10^{34}$ erg s$^{-1}$) and
at least for the OG model fΩ is expect to drop well below 1. However, with the
wide range of gamma-ray efficiencies in the plot it is not possible to discriminate
between these model predictions. Finally for $\dot E <10^{34}$ erg s$^{-1}$ the sample is
dominated by the MSPs and the trend seems more consistent with $\dot E$.

\section{DISCUSSION}
The LAT is providing a new local population of gamma-ray pulsars, composed by tree families: the radio loud normal pulsars, the gamma-selected pulsars, and the millisecond pulsars. Considering also that gamma-ray pulsations have been searched for pulsars spanning the entire $P-\dot P$ plane, it makes of the LAT population an unbiased one that provide a new window on pulsar demographics and physics. 

Evidence is that a large fraction of the local energetic pulsars are GeV
emitters. There is also a significant correlation with X-ray and TeV bright pulsar
wind nebulae. Conversely, the pulsar origin of a large fraction 
of the bright unidentified Galactic EGRET sources is now uncovered. Plausible
pulsar counterparts are also found for several previously detected TeV sources. In this sense
the ``mystery'' of the unidentified EGRET sources is largely solved.

Finally, the light curve and spectral evidence summarized above suggests that these
pulsars have high altitude emission zones whose fan-like beams scan over a large portion
of the celestial sphere. This is also confirmed by the estimated pulsars birth rate and by the ratio of radio-selected to gamma-selected gamma-ray pulsars.

\acknowledgments
The \textit{Fermi} LAT Collaboration acknowledges the generous
support of a number of agencies and institutes that
have supported the Fermi LAT Collaboration. These
include the National Aeronautics and Space Administration
and the Department of Energy in the United States,
the Commissariat \`a l'Energie Atomique and the Centre
National de la Recherche Scientifique / Institut National
de Physique Nucl\'eaire et de Physique des Particules
in France, the Agenzia Spaziale Italiana and the Istituto
Nazionale di Fisica Nucleare in Italy, the Ministry
of Education, Culture, Sports, Science and Technology
(MEXT), High Energy Accelerator Research Organization
(KEK) and Japan Aerospace Exploration Agency
(JAXA) in Japan, and the K. A. Wallenberg Foundation
and the Swedish National Space Board in Sweden.

Additional support for science analysis during the operations phase is gratefully
acknowledged from the Istituto Nazionale di Astrofisica in Italy and the 
Centre National d'\'Etudes Spatiales in France.

The Parkes radio telescope is part of the Australia Telescope which is funded by the Commonwealth
Government for operation as a National Facility managed by CSIRO.  The Green Bank Telescope is operated
by the National Radio Astronomy Observatory, a facility of the National Science Foundation operated
under cooperative agreement by Associated Universities, Inc. The Arecibo Observatory is part of the
National Astronomy and Ionosphere Center (NAIC), a national research center operated by Cornell
University under a cooperative agreement with the National Science Foundation. The Nan\c cay Radio
Observatory is operated by the Paris Observatory, associated with the French Centre National de la
Recherche Scientifique (CNRS). The Lovell Telescope is owned and operated by the University of
Manchester as part of the Jodrell Bank Centre for Astrophysics with support from the Science and
Technology Facilities Council of the United Kingdom. The Westerbork Synthesis Radio Telescope is
operated by Netherlands Foundation for Radio Astronomy, ASTRON.

\bigskip 
\bibliography{CaliandroFermiS}

\end{document}